\documentclass[a4paper]{article}
\pdfoutput=1

\usepackage{graphicx}
\usepackage{amsmath,amssymb,amsfonts}
\usepackage{fullpage}
\usepackage{algorithm}
\usepackage{algpseudocode}
\algrenewcommand{\algorithmiccomment}[1]{//#1}
\usepackage{textcomp}
\usepackage{multirow}
\usepackage[hidelinks]{hyperref}
\usepackage{authblk}

\newcommand{\edit}{\textrm{edit}}

\title{On the Maximal Independent Sets of $k$-mers with the Edit Distance}
\author[1,2]{Leran Ma}
\author[2]{Ke Chen}
\author[2,3]{Mingfu Shao}
\affil[1]{\footnotesize Schreyer Honors College, The Pennsylvania State University, United States}
\affil[2]{\footnotesize Department of Computer Science and Engineering, The Pennsylvania State University, United States}
\affil[3]{\footnotesize Huck Institutes of the Life Sciences, The Pennsylvania State University, United States}

\begin{document}
\maketitle

\begin{abstract}
In computational biology, $k$-mers and edit distance are
fundamental concepts. 
However, little is known 
about the metric space of all $k$-mers equipped with the edit distance.
In this work, we explore the structure of the
$k$-mer space by studying its maximal independent sets (MISs).
An MIS is a sparse sketch of all $k$-mers with nice theoretical properties,
and therefore admits critical applications in clustering, indexing, hashing, and sketching 
large-scale sequencing data, particularly those with high error-rates. 
Finding an MIS is a challenging problem, as the size of a $k$-mer space grows geometrically
with respect to $k$. 
We propose three algorithms for this problem.
The first and the most intuitive one uses a greedy strategy. 
The second method implements two techniques to avoid redundant
comparisons by taking advantage of the locality-property of
the $k$-mer space and the estimated bounds on the edit distance. The
last algorithm avoids expensive calculations of the edit distance by
translating the edit distance into the shortest path in a specifically designed graph. 
These algorithms are implemented and the calculated MISs of $k$-mer spaces and their 
statistical properties are reported and analyzed
for $k$ up to 15.  Source code is freely available at {https://github.com/Shao-Group/kmerspace}.

\end{abstract}

\section{Introduction}

Given an alphabet $\Sigma$, we denote by $S_k$ the set of all sequences
of length $k$ over $\Sigma$. Clearly, $|S_k| = |\Sigma|^k$.
In this work, we call a sequence of length $k$ a $k$-mer, and
call $S_k$ the $k$-mer space, for the sake of convenience.
The Levenshtein distance~\cite{editdistance}, also known as the edit distance, between two sequences $u$ and $v$,
denoted as $\edit(u,v)$,
is defined as the minimum number of insertions, deletions, and substitutions
needed to transform $u$ into $v$. 
The edit distance is a metric;
the $k$-mer space together with the edit distance $(S_k, \edit)$ forms a metric space.

Despite the ubiquitous use of $k$-mers and the edit distance in computational biology,
the properties and structures of the metric space $(S_k,\edit)$ remain largely unknown.
It is worth noting that $(S_k,\edit)$ is intrinsically different from the well-studied metric spaces on sequences such as
the Hamming distance space or other normed spaces~(when a $k$-mer is represented as a vector of dimension $k$), as evidenced by the fact that
embedding the edit distance on $\{0,1\}^k$ into the $L_1$ space requires a distortion of $\Omega(\log k)$~\cite{embedding2, KR09}.

In this work, 
we study the \emph{maximal independent set}~(MIS) of a $k$-mer space parameterized with an integer $d \ge 0$.
Formally, we say a subset $M\subset S_k$ is \emph{independent} if $\edit(u,v) > d$ for every two $k$-mers $u$ and $v$ in $M$,
and we say an independent subset $M$ is \emph{maximal} if there does not exist another independent subset $M'$ such that $M\subsetneq M'$.
These definitions can be equivalently stated using the language of graph. 
Given an integer $d \ge 0$, we define an undirected graph $G^d_k = (V_k = S_k, E^d_k)$ where there is an edge
$(u,v)\in E_k^d$ if $\edit(u,v) \le d$.
The above defined MISs of the $k$-mer space with respect to an integer $d$ is equivalent to MISs in the graph $G^d_k$.

We address the problem of finding an MIS of a $k$-mer space.
Solving this problem helps understand the structure of the metric space $(S_k,\edit)$ as
an MIS indicates how dense a $k$-mer space is under the edit distance.
More importantly, finding MISs of $G^d_k$ has critical applications in large-scale sequence
analysis thanks to its nice properties.
We elaborate on a few below.

First, an MIS $M$ naturally can be used as a set of ``centers'' for clustering $k$-mers.
The centers~($k$-mers in an $M$) are distant from each other which means that clusters are not crammed together.
In addition, every $k$-mer $v$ can find a ``nearby'' center, as there exists a $k$-mer $u\in M$
such that $\edit(u,v) \le d$~(otherwise $M\cup \{v\}$ is an independent set
that strictly contains $M$).
These properties make an MIS an ideal set of centers for clustering $k$-mers.

Second, $k$-mers of an MIS can be used to design a new seeding scheme for aligning error-prone sequencing reads.
Modern fast aligners often use exact $k$-mer matches as anchors~(or seeds)~\cite{star, minimap2}.
However, such methods exhibits low sensitivity when aligning reads with high error-rate~(i.e., long-reads data
generated by PacBio~\cite{rhoads2015pacbio} and Oxford Nanopore~\cite{jain2018nanopore} protocols),
because under high error-rate, two biologically related sequences hardly 
share any identical $k$-mers (for a reasonable choice of $k$ such as $k = 15$ and an error rate of 15\%).
One can choose to use smaller $k$ but this will lead to high false 
positives as unrelated sequences can, by chance, share many short $k$-mers.
A new seeding scheme can be designed by first mapping a $k$-mer into its 
nearest $k$-mer $u$ in an MIS $M$: two $k$-mers form an anchor if they are mapped
to the same $k$-mer in $M$. The advantage of this MIS-based seeding scheme
is that it tolerates edits in $k$-mers and is therefore more sensitive for
error-prone sequencing data.  Comparing to other $k$-mer-alternative seeding methods such as
spaced seeds~\cite{flash, patternhunter} and indel seeds~\cite{mak2006indel}
where only restricted types of edits are allowed~(for example, spaced seeds cannot model indels),
this MIS-based scheme can recognize
similar $k$-mers with both indels and substitutions.
Moreover, the intrinsic properties of an MIS also control the false-positives:
by triangle inequality,
two distant $k$-mers that are $2d$ edits apart will never be
mapped to the same $k$-mer.

Third, an MIS of the $k$-mer space can be used to improve \emph{sketching} approaches.
Sketching enables scaling to large datasets by only selecting a subset of representative $k$-mers in a sequence.
Existing $k$-mer sketching methods such as FracMinHash~\cite{sourmash, fracminhash} often 
use a random permutation of $k$-mers and pick the top fraction
as representative $k$-mers. By using an MIS as a representative subset instead,
the selected $k$-mers are guaranteed to be at least of $d$ edits apart of each other,
and are hence more efficient.

All above applications require finding an MIS in the first place.
However, computing an MIS of a $k$-mer space is difficult, 
simply because the number of $k$-mers $|S_k| = |\Sigma|^k$
(equivalently, the number of vertices in the graph $G_k^d$) grows
exponentially with $k$ and the number of edges in the graph
$|E_k^d| = O(|\Sigma|^{2k})$ grows even faster. 
Existing algorithms for finding an MIS
in a general graph, for example, the greedy algorithm that takes $O(m)$ time
for a graph with $m$ edges, are therefore not suitable for our problem.
In fact, even constructing $E_k^d$ explicitly is often unaffordable.
In this work, we design three algorithms for finding an MIS of a $k$-mer space.
Our algorithms take into account the special properties of $k$-mers and the edit distance so that the underlying graph $G_k^d$ does not need to be explicitly
constructed.
Our algorithms are able to scale to instances with $k = 15$.
We implemented these algorithms and calculated
an MIS for all combinations of $k$ and $d$ for $k \le 15$ and reported their statistics.
We also analyzed and concluded when to use which algorithm for different combinations of $k$ and $d$.



\section{Algorithms}

We design three algorithms for finding a MIS of all $k$-mers, given $k$ and $d$.
The first one is a greedy
algorithm that is similar to the greedy algorithm for computing an MIS on general graphs
but without explicitly building all edges of the underlying graph.
The second algorithm improves the first one by 
reducing redundant comparisons through recognizing the locality properties of $k$-mers
and estimating and incorporating the bounds of the edit distance into the algorithm. The last algorithm
transforms calculating the edit distance into finding the shortest path
in a specifically constructed graph, and finding an MIS is then transformed into efficient graph traversing
together with accompanied data structures to speed up.

\subsection{A Simple Greedy Algorithm}
This algorithm maintains a (dynamic) array $M$ that stores the current
MIS, initialized as an empty array. The algorithm 
examines each $k$-mer in $S_k$: for the current $k$-mer $v$,
it compares with all $k$-mers in $M$, and adds $v$ to $M$
if $\edit(u,v) > d$ for every $u\in M$.
See the pseudocode given below.

We now show that this algorithm is correct, i.e., the returned $M$ is indeed an MIS. 
According to the algorithm, $k$-mer $v$ is added to $M$ only if $\edit(u, v) > d$ for all $u\in M$.
This means that any two $k$-mers in $M$ have an edit distance larger than $d$, i.e., $M$ is independent.
On the other hand, for any $k$-mer $v$ not included in $M$, the algorithm guarantees that there exists
a $k$-mer $u$ in $M$ such that $\edit(u,v) \le d$; 
this implies that $M$ is maximal.

\begin{algorithm}[H]
  \caption{Simple Greedy Algorithm}\label{alg:cap}
  \begin{algorithmic}
    \State initialize an empty array $M$
    \For{each $k$-mer $v \in S_k$}
    \State $\mathit{isMapped} \gets \text{false}$
    \For{each $k$-mer $u\in M$}
    \If{$\edit(u, v)\leq d$}
    \State $\mathit{isMapped} \gets \text{true}$
    \State break
    \EndIf
    \EndFor
    \If{$\mathit{isMapped} == \text{false}$}
    \State add $v$ into $M$
    \EndIf
    \EndFor
    \State return $M$
  \end{algorithmic}
\end{algorithm}

The above algorithm runs in $O(|M|\cdot |\Sigma|^k\cdot d\cdot k)$ time,
as in the worst case it compares each $k$-mer in $S_k$ with all $k$-mers in $M$,
and determining if $\edit(u,v)\le d$ for two $k$-mers $u$ and $v$ takes $O(d\cdot k)$ time.
The running time is output-sensitive.
It is in favor of instances with small MIS.

\subsection{An Improved Greedy Algorithm}
We say a $k$-mer $v$ is \emph{mapped} to a $k$-mer $u$ in the (partially) constructed MIS $M$
if $\edit(u, v)\leq d$.
Such a mapping explains why $v$ is not selected into the MIS.
Algorithm~\ref{alg:cap} essentially finds a mapping for each $k$-mer $v\in S_k\setminus M$
using an iterative, exhaustive search.
As the size of an MIS increases, this can significantly impair the
performance of the algorithm.
Observe that the search order matters: if a mapping of a $k$-mer is found early, all the
following comparisons can be avoided.
Furthermore, based on 
the results of previous
comparisons, it can be inferred that some $k$-mers in the MIS are close to $v$
while some others are too far away to serve as mappings.
We design two techniques that allow us to quickly determine if $v$ can be mapped
to some $k$-mer in $M$, by utilizing the locality-properties,
and to quickly filter out those $k$-mers in $M$ that $v$ cannot be mapped to,
by estimating the bounds on the edit distance.
These two techniques serve as extensions to the first algorithm to avoid unnecessary comparisons.

The first technique is based on the observation that a $k$-mer is likely to have
a shared mapping with its ``neighboring'' $k$-mers.
For a $k$-mer $v$, we 
define its \emph{nearest neighbors} to be the subset of $S_k$ in which each $k$-mer has edit distance $1$ from $v$, i.e., a substitution. 
For any given $k$-mer, clearly, its nearest neighbor set contains $k(|\Sigma|-1)$ $k$-mers.
For example, the set of nearest neighbors of the 3-mer $AAA$ is
$\{AAC,AAG,AAT,ACA,AGA,ATA,CAA,GAA,TAA\}$.
If $AAA$ is selected into the MIS, all the above 9 $k$-mers will have it as a shared mapping.
This locality-property suggests a modification to Algorithm 1: 
store the mapping information of $k$-mers (once found) so that potential shared mappings can be examined first.
Note that a $k$-mer can have multiple valid mappings in the MIS, but we only keep record of one to save both time and space.
When checking whether a $k$-mer $v$ should be added to the MIS or not, we first
compute the nearest neighbors of $v$ and check if $v$ shares a mapping with them.
If a shared mapping is found, we store this mapping information for $v$ and directly conclude that $v$ cannot be included in the MIS.




\begin{algorithm}[!b]
  \caption{Improved Greedy Algorithm}\label{alg:heuristic}
  \begin{algorithmic}
    \State initialize empty arrays $M$, $mapping$, and $d\sigma$ \Comment{$d\sigma[u]$ is later used to store $\edit (u, \sigma^k)$, for each $\sigma\in\Sigma$}
    \For{each $k$-mer $v \in S_k$}
    \State $\mathit{isMapped} \gets \text{false}$
    \For{each nearest neighbor $u$ of $v$}
    \If{$\edit\left(mapping[u], v\right)\leq d$}
    \State $\mathit{isMapped} \gets \text{true}$
    \State $mapping[v] \gets mapping[u]$
    \State break
    \EndIf
    \EndFor
    \If{$\mathit{isMapped} == \text{true}$}
    \State continue
    \EndIf
    \State $v_{\sigma} \gets \edit\left(v, \sigma^k\right)$, $\forall \sigma\in\Sigma$
    \For{each $k$-mer $u \in M$}
    \If{$\max_{\sigma\in\Sigma}\{|v_{\sigma}-d\sigma[u]|\}>d$}
    \State continue
    \ElsIf{$\min_{\sigma\in\Sigma}\{v_{\sigma}+d\sigma[u]\}\leq d$ or $\edit(u, v)\leq d$}
    \State $mapping[v] \gets u$
    \State $\mathit{isMapped} \gets \text{true}$
    \State break
    \EndIf
    \EndFor
    \If{$\mathit{isMapped} == \text{false}$}
    \State add $v$ into $M$
    \State $mapping[v] \gets v$
    \State $d\sigma[v] \gets v_\sigma$, $\forall\sigma\in\Sigma$
    \EndIf
    \EndFor
    \State output $M$
  \end{algorithmic}
\end{algorithm}

The second technique uses the estimated bounds of the edit distance.
Recall that $(S_k, \edit)$ is a metric space, in particular,
the triangle inequality holds.
Let $\sigma^k$ be the $k$-mer that is purely composed of the character
$\sigma$. 
For any two $k$-mers $u, v\in S_k$ and $\sigma\in\Sigma$, we have
\begin{align}
\edit(u, v) &\geq \left|\edit\left(u, \sigma^k\right) - \edit\left(\sigma^k, v\right)\right|,\label{eq:triangle}\\
\edit(u, v) &\leq \edit\left(u, \sigma^k\right) + \edit\left(\sigma^k, v\right).\label{eq:triangle2}
\end{align}
Note that the calculation of $\edit(u, \sigma^k)$ is simple:
all characters in $u$ that are not $\sigma$ should be substituted by $\sigma$
(insertions and deletions cannot reduce the number of edits needed).
Hence, the upper and lower bounds provided by inequalities~\eqref{eq:triangle} and \eqref{eq:triangle2}
can be used as a filter before performing the expensive calculation of $\edit(u, v)$.
Using this idea, we modify the algorithm as follows.
For each $k$-mer $u$ in the constructed MIS, we store additional values
$\edit(u, \sigma^k)$ for each $\sigma\in\Sigma$.
When searching for a mapping of a $k$-mer $v$, for each $k$-mer $u$ in the MIS,
we check if the parameter $d$ is within the range
given by inequalities~\eqref{eq:triangle} and \eqref{eq:triangle2}.
If so, we calculate the exact distance $\edit(u,v)$ and compare it with $d$;
otherwise, if $d$ is less than
$\max_{\sigma\in\Sigma}\left\{\left|\edit\left(u,\sigma^k\right)
-\edit\left(\sigma^k,v\right)\right|\right\}$,
we directly conclude that $\edit(u,v)>d$ and therefore $u$ cannot be a valid
mapping of $v$;
if $d$ is greater than
$\min_{\sigma\in\Sigma} \left\{\edit\left(u,\sigma^k\right)
+\edit\left(\sigma^k,v\right)\right\}$,
then we know that $\edit(u,v)<d$ and $v$ cannot be included in the MIS.


Algorithm~\ref{alg:heuristic} incorporates above two techniques.
Note that Algorithm~\ref{alg:heuristic} is also correct, i.e., the returned
$M$ is guaranteed an MIS, following the correctness of Algorithm~1 and above analysis.
The worst-case running time of Algorithm~\ref{alg:heuristic}
is the same with Algorithm~1 but it runs faster in practice~(see Section~3).

\subsection{A BFS-based Algorithm}

In the third algorithm, we build a graph in which vertices correspond to
all $k$-mers and ($k-1$)-mers, and edges correspond to pairs of vertices
with edit distance being exactly 1 except that there is no edge between any two $(k-1)$-mers.
For example, if we take $k=4$, the graph contains all the 4-mers and 3-mers.
There is no edge between any pair of 3-mers.
Two 4-mers are connected if one can be obtained from the other by
a single substitution.
A 4-mer and a 3-mer are connected if the 4-mer can be obtained by
inserting a character to the 3-mer.

The key property of this graph is that two $k$-mers $u$ and $v$
have $\edit(u,v) = d$ if and only if the distance~(the length
of the shortest path) between $u$ and $v$ in this graph is $d$.
To formally see this, by the construction of the graph, every path between two $k$-mers gives a valid
sequence of edits transforming one to the other.
Hence, we only need to show that given a sequence of $d$ edits that transforms
a $k$-mer $u$ to a $k$-mer $v$, the edits can be rearranged such that it
corresponds to a path of length $d$ from $u$ to $v$ in the graph.
The only issue comes from the fact that the given sequence of edits may
contain several insertions (or deletions) in a row so the intermediate strings
may not be $k$-mers or $(k-1)$-mers.
However, because both $u$ and $v$ are $k$-mers, the sequence of edits must contain
the same number of insertions and deletions.
Thus the sequence can be rearranged so that each deletion is followed immediately
by an insertion.
Such a sequence has a direct representation in the graph.
For example, the transformation from the 5-mer $TGATT$ to the 5-mer $ATTGA$ can be represented by the following shortest path of length 4:
$TGATT \rightarrow GATT \rightarrow GATTG \rightarrow ATTG \rightarrow ATTGA$.

Following the above property,
the calculation of all $k$-mers within an edit distance
of at most $d$ from a $k$-mer $u$ can be (efficiently) achieved
by \emph{exploring} $u$ using breadth-first search, i.e., traversing all vertices reachable
from $u$ within a distance of $d$.  Again, if $u$ is already added to the MIS, then all vertices found during the exploration can be marked as mapped to $u$.

We do not need to fully explore every single $k$-mer.
Instead, we can reuse the information stored in exploring previous vertices to stop early.
Specifically, for each vertex $u$ (not in the current MIS), we store the 
distance from $u$ to any vertex in the MIS, and update it in the exploring.
See Algorithm~\ref{alg:bfs} for the complete pseudocode. 

\begin{algorithm}[H]
  \caption{BFS-based Algorithm}\label{alg:bfs}
  \def\distance{\mathit{distance}}
  \def\frontier{\mathit{frontier}}
  \begin{algorithmic}
    \State initialize empty arrays $\distance$ and $M$
    \State $\distance[v] \gets \infty$ for all $k$-mers $v$
    \State $\distance[u] \gets \infty$ for all ($k-1$)-mers $u$
    \For{each $k$-mer $v \in S_k$}
    \If{$\distance[v]==\infty$}
    \Comment{explore $v$}
    \State add $v$ to $M$ 
    \State $\distance[v]\gets 0$
    \State initialize an empty queue $\frontier$
    \State $\frontier.$add$(v)$
    \While{$\frontier$ is not empty}
    \State $u \gets \frontier.$pop$()$
    \For{each adjacent vertex $w$ of $u$ in the graph}
    \If{$\distance[w]>\distance[u]+1$}
    \State $\distance[w]\gets \distance[u]+1$
    \If{$\distance[w]<d$}
    \State $\frontier.$add$(w)$
    \EndIf
    \EndIf
    \EndFor
    \EndWhile
    \EndIf
    \EndFor
    \State return $M$
  \end{algorithmic}
\end{algorithm}


The exploring step~(variant of BFS) guarantees that all the unexplored $k$-mers must have an edit distance greater than $d$ from all vertices in the MIS, and all the explored $k$-mers must have an edit distance less than or equal to $d$ from a vertex in the MIS.
Hence, $k$-mers in the returned MIS are at least $d+1$ edits apart (i.e., independent)
and no other $k$-mers can be added to the resulting MIS (i.e., maximal).
This shows that Algorithm~3 is correct.

To see its time complexity, note that the initial distance of a vertex is at most $d$, hence it can
be updated (decreased) at most $d-1$ times by the subsequent explorings. 
In other words, each vertex can be explored at most $d$ times.
Each time a $k$-mer is explored, we compute its neighbors with distance $1$.
The number of neighbors increases linearly with respect to $k$ (for a constant-size alphabet).
Thus, the running time for this algorithm is $O(d\cdot k\cdot |V|)$, where $|V|=|\Sigma|^k+|\Sigma|^{k-1}$ is the total number of $k$-mers and  ($k-1$)-mers.
The space complexity of this algorithm is $O(|V|)$.
For large values of $k$, the graph is dense and too large to be stored in memory, we 
compute edges of the graph on the fly.
The space is mainly used to store the distance array, which grows linearly with respect to the number of vertices.

\section{Results}

\begin{figure*}[!htb]
\includegraphics[width=\textwidth]{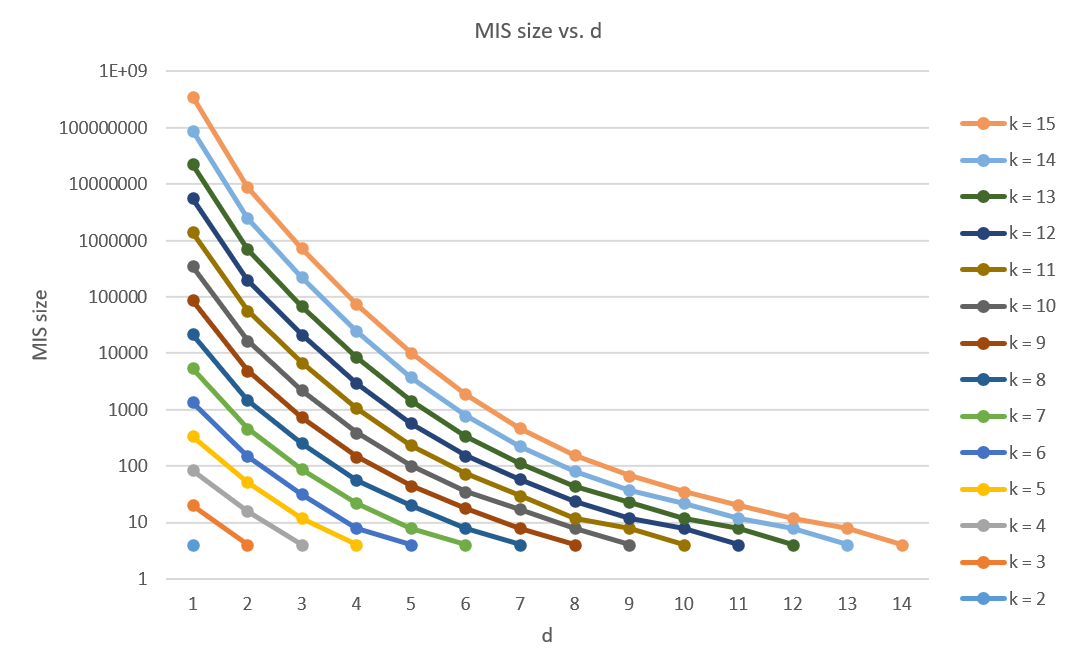}
\centering
\caption{The size of an MIS~(found by the fastest algorithm) with respect to parameters $k$ and $d$.} \label{fig1}
\end{figure*}

We implemented all three algorithms described above
and conducted experiments for $2 \le k \le 15$ and $d < k$
(with the DNA alphabet $\Sigma=\{A, C, G, T\}$). 
Table~\ref{tab1} reports the sizes of the resulting MISs
which are depicted in Fig.~\ref{fig1}.
For each row (i.e., with the same $d$ value), the size of the MIS increases
(approximately) geometrically with respect to $k$, which is consistent with the
geometric growth of the size of the $k$-mer space $S_k$.  For each column
(i.e., with the same $k$ value), the size of the MIS decreases geometrically
with respect to the growth of $d$. As discussed before, $d$ specifies the
sparsity of the resulting MIS. With larger $d$ value, fewer vertices can be
selected into the MIS.

Table~\ref{tab2} reports the
fastest algorithm to calculate the MIS for each combination of $k$ and $d$.
Table~\ref{tab2} together with Table~\ref{tab1} suggests a general strategy to
choose the best algorithm for a specific combination of $k$ and $d$.  
Algorithm 1 is efficient for $d$ values within the interval
$[k-4,k-1]$ since these $d$ values result in an MIS with less than 15 vertices
approximately.  The BFS-based Algorithm~(Algorithm 3) is the best for $d$ values between 1 and 4
because such a small $d$ leads to an MIS that is too large to perform pairwise
comparisons efficiently.  The rest of the $d$ values are better be handled by Algorithm~2.

Table~\ref{tab3} and Table~\ref{tab4} record the corresponding time and memory
usage, respectively.  The rows of Table~\ref{tab3} show that the running time
generally increases geometrically with respect to the parameter $k$. The
current computation bottleneck is the case with $k=15$ and $d=5$, which takes
approximately 15 hours using Algorithm~2.
Algorithm 2 always has the largest memory overhead comparing with
other two. The extra memory is mainly used to store the mapping
information of all $k$-mers. The peak memory usage is
approximately 2 GB for $k=15$ and $d\in\{5, 6, \ldots, 10\}$.

\begin{table*}[!ht]
\caption{The size of calculated MIS~(by the fastest algorithm) for different $k$ and $d$.}\label{tab1}
\centering
\resizebox{\textwidth}{!}{
\begin{tabular}{p{5pt}rrrrrrrrrrrrrrrrrrrrrrr}
\multicolumn{2}{}{} & \multicolumn{14}{c}{$k$} \\
&     & 2 & 3 & 4 & 5 & 6 & 7 & 8 & 9 & 10     &  11     & 12      & 13      & 14      & 15   \\ \cline{2-16}
\multirow{14}{*}{$d$}
&  1  & 4 & 16 & 64 & 256 & 1024 & 4096 & 16384 & 65536 & 262144 & 1048576 & 4194304 & 16777216 & 67108864 & 268435456 \\
&  2  &  & 4 & 12 & 36 & 96 & 311 & 1025 & 3451 & 11743 & 40604 & 141943 & 500882 & 1782677 & 6388106 \\
&  3  &  &  & 4 & 8 & 20 & 57 & 164 & 481 & 1463 & 4574 & 14522 & 46908 & 153767 & 510118 \\
&  4  &  &  &  & 4 & 4 & 14 & 34 & 90 & 242 & 668 & 1894 & 5517 & 16440 & 49992 \\
&  5  &  &  &  &  & 4 & 4 & 12 & 25 & 57 & 133 & 338 & 879 & 2346 & 6486 \\
&  6  &  &  &  &  &  & 4 & 4 & 10 & 17 & 38 & 79 & 188 & 448 & 1107 \\
&  7  &  &  &  &  &  &  & 4 & 4 & 9 & 13 & 28 & 54 & 112 & 251 \\
&  8  &  &  &  &  &  &  &  & 4 & 4 & 4 & 12 & 20 & 37 & 75 \\
&  9  &  &  &  &  &  &  &  &  & 4 & 4 & 4 & 11 & 14 & 30 \\
& 10  &  &  &  &  &  &  &  &  &  & 4 & 4 & 4 & 10 & 13 \\
& 11  &  &  &  &  &  &  &  &  &  &  & 4 & 4 & 4 & 8 \\
& 12  &  &  &  &  &  &  &  &  &  &  &  & 4 & 4 & 4 \\
& 13  &  &  &  &  &  &  &  &  &  &  &  &  & 4 & 4 \\
& 14  &  &  &  &  &  &  &  &  &  &  &  &  &  & 4 
\end{tabular}
}
\end{table*}

\begin{table*}[!htb]
  \caption{The fastest algorithm for each combination of $k$ and $d$. Numbers 1, 2, and 3 correspond to Algorithms~1, 2, and 3.
  }\label{tab2}
\centering
\begin{tabular}{p{5pt}rrrrrrrrrrrrrrrrrrrrrrrr}
\multicolumn{2}{}{} & \multicolumn{14}{c}{$k$} \\
&     & 2 & 3 & 4 & 5 & 6 & 7 & 8 & 9 & 10     &  11     & 12      & 13      & 14      & 15   \\ \cline{2-16}
\multirow{14}{*}{$d$}
&  1  & 1 & 1 & 1 & 1 & 1 & 3 & 3 & 3 & 3 & 3 & 3 & 3 & 3 & 3 \\
&  2  &  & 1 & 1 & 1 & 1 & 1 & 3 & 3 & 3 & 3 & 3 & 3 & 3 & 3 \\
&  3  &  &  & 1 & 1 & 1 & 1 & 2 & 3 & 3 & 3 & 3 & 3 & 3 & 3 \\
&  4  &  &  &  & 1 & 1 & 1 & 1 & 2 & 2 & 2 & 2 & 3 & 3 & 3 \\
&  5  &  &  &  &  & 1 & 1 & 1 & 2 & 2 & 2 & 2 & 2 & 2 & 2 \\
&  6  &  &  &  &  &  & 1 & 1 & 1 & 1 & 2 & 2 & 2 & 2 & 2 \\
&  7  &  &  &  &  &  &  & 1 & 1 & 1 & 1 & 2 & 2 & 2 & 2 \\
&  8  &  &  &  &  &  &  &  & 1 & 1 & 1 & 1 & 1 & 2 & 2 \\
&  9  &  &  &  &  &  &  &  &  & 1 & 1 & 1 & 1 & 2 & 2 \\
& 10  &  &  &  &  &  &  &  &  &  & 1 & 1 & 1 & 1 & 2 \\
& 11  &  &  &  &  &  &  &  &  &  &  & 1 & 1 & 1 & 1 \\
& 12  &  &  &  &  &  &  &  &  &  &  &  & 1 & 1 & 1 \\
& 13  &  &  &  &  &  &  &  &  &  &  &  &  & 1 & 1 \\
& 14  &  &  &  &  &  &  &  &  &  &  &  &  &  & 1 
\end{tabular}
\end{table*}

\begin{table*}[!htb]
  \caption{Running time (seconds) of the fastest algorithm for each $(k,d)$.}\label{tab3}
\centering
\begin{tabular}{p{5pt}rrrrrrrrrrrrrrrrrrrrrrr}
\multicolumn{2}{}{} & \multicolumn{14}{c}{$k$} \\
&     & 2 & 3 & 4 & 5 & 6 & 7 & 8 & 9 & 10     &  11     & 12      & 13      & 14      & 15   \\ \cline{2-16}
\multirow{14}{*}{$d$}
&  1  & 0 & 0 & 0 & 0 & 0 & 0 & 0 & 1 & 5 & 25 & 114 & 528 & 3382 & 14028 \\
&  2  &  & 0 & 0 & 0 & 0 & 0 & 0 & 2 & 11 & 46 & 202 & 977 & 5444 & 22380 \\
&  3  &  &  & 0 & 0 & 0 & 0 & 0 & 3 & 18 & 78 & 344 & 2161 & 10317 & 46701 \\
&  4  &  &  &  & 0 & 0 & 0 & 0 & 1 & 8 & 55 & 460 & 4191 & 21801 & 36614 \\
&  5  &  &  &  &  & 0 & 0 & 0 & 0 & 4 & 25 & 152 & 1409 & 11887 & 53478 \\
&  6  &  &  &  &  &  & 0 & 0 & 0 & 3 & 17 & 94 & 659 & 4423 & 23922 \\
&  7  &  &  &  &  &  &  & 0 & 0 & 1 & 13 & 77 & 504 & 3008 & 14864 \\
&  8  &  &  &  &  &  &  &  & 0 & 1 & 7 & 52 & 438 & 2422 & 11308 \\
&  9  &  &  &  &  &  &  &  &  & 0 & 5 & 31 & 258 & 2128 & 10003 \\
& 10  &  &  &  &  &  &  &  &  &  & 4 & 23 & 210 & 1809 & 9197 \\
& 11  &  &  &  &  &  &  &  &  &  &  & 21 & 187 & 1225 & 7322 \\
& 12  &  &  &  &  &  &  &  &  &  &  &  & 184 & 988 & 5548 \\
& 13  &  &  &  &  &  &  &  &  &  &  &  &  & 931 & 4743 \\
& 14  &  &  &  &  &  &  &  &  &  &  &  &  &  & 4332
\end{tabular}
\end{table*}

\begin{table*}[!htp]
\caption{Memory usage (kB) of the fastest algorithm for each $(k,d)$.}\label{tab4}
\centering
\resizebox{\textwidth}{!}{
\begin{tabular}{p{5pt}rrrrrrrrrrrrrrrrrrrrrr}
\multicolumn{2}{}{} & \multicolumn{14}{c}{$k$} \\
&     & 2 & 3 & 4 & 5 & 6 & 7 & 8 & 9 & 10     &  11     & 12      & 13      & 14      & 15   \\ \cline{2-16}
\multirow{14}{*}{$d$}
&  1  & 1076 & 1076 & 1076 & 1076 & 1336 & 1332 & 1328 & 1324 & 3648 & 4568 & 8524 & 23916 & 85316 & 331164 \\
&  2  &  & 1076 & 1076 & 1076 & 1076 & 1076 & 1324 & 1320 & 3544 & 4596 & 8364 & 24012 & 85396 & 331220 \\
&  3  &  &  & 1076 & 1076 & 1076 & 1072 & 1476 & 1532 & 4008 & 5148 & 8888 & 24336 & 85984 & 331908 \\
&  4  &  &  &  & 1076 & 1076 & 1076 & 1076 & 1624 & 5460 & 11508 & 36224 & 29512 & 92524 & 341752 \\
&  5  &  &  &  &  & 1076 & 1072 & 1076 & 1648 & 5352 & 11468 & 36016 & 134416 & 527728 & 2100852 \\
&  6  &  &  &  &  &  & 1072 & 1076 & 1072 & 1952 & 11584 & 36272 & 134484 & 527672 & 2100688 \\
&  7  &  &  &  &  &  &  & 1076 & 1076 & 1880 & 2012 & 36164 & 134480 & 527688 & 2100568 \\
&  8  &  &  &  &  &  &  &  & 1072 & 1920 & 1868 & 1952 & 1920 & 527772 & 2100568 \\
&  9  &  &  &  &  &  &  &  &  & 1980 & 2012 & 3464 & 1920 & 527760 & 2100548 \\
& 10  &  &  &  &  &  &  &  &  &  & 1980 & 1868 & 3456 & 1964 & 2100636 \\
& 11  &  &  &  &  &  &  &  &  &  &  & 2032 & 3396 & 3184 & 3308 \\
& 12  &  &  &  &  &  &  &  &  &  &  &  & 1976 & 3380 & 1992 \\
& 13  &  &  &  &  &  &  &  &  &  &  &  &  & 3404 & 1964 \\
& 14  &  &  &  &  &  &  &  &  &  &  &  &  &  & 3384
\end{tabular}
}
\end{table*}

\section{Conclusion and Discussion}

We studied the problem of extracting an MIS as a representative substructure from
a $k$-mer space. Three algorithms are designed to
efficiently solve the problem for different ranges of parameters $k$ and $d$.
The first one is a simple greedy algorithm similar to the greedy graph for general graphs. The
second algorithm extends the first by implementing two techniques to avoid some
redundant comparisons. The third algorithm represents the edit distance as a
shortest path in an extended graph and uses a variant of BFS. Experiments are done for
$k$ up to 15. The computation bottleneck occurs at $k=15$ and
$d=5$ where the second algorithm performs the best. The corresponding peak
running time is approximately 15 hours, and the peak memory usage is about 2
GB.

For future work, we would like to extend this study to larger $k$-mer spaces.
Considering the current computation bottleneck, one potential improvement is to design more 
heuristics to partition the $k$-mer space so that only a small subset of $k$-mers will be involved in the expensive pairwise edit distance computation.

Another idea is to revise the BFS-based algorithm.
The extended graph with both $k$-mers and $(k-1)$-mers is highly symmetric
with respect to permutations of the alphabet.
For example, the search tree for the 5-mers $AGAAC$ and $TATTG$
are isomorphic where $A$, $G$, and $C$ are replaced with $T$, $A$, and $G$, respectively.
This algorithm may take advantage of such symmetries to explore the graph more efficiently.
Identifying and utilizing symmetric substructures of the $k$-mer space is of independent interest and has applications in other related fields~\cite{dfa}.

Regarding the practical applications of the constructed MISs, one potentiality is their usage in $k$-mer-based reads clustering problems~\cite{cluster1,cluster2},
where the choice of centers is crucial.
Because an MIS contains a group of representatives for the $k$-mer space that are guaranteed to be a certain edit distance apart,
it is ideal for generating even-sized clusters.
Moreover, the MISs can also be used to generate keys for $k$-mer-based reads
hashing problems~\cite{hash1,hash2} where nearby $k$-mers are desired to share
a hash value~(i.e., locality-sensitive hashing), which is particularly useful
in aligning error-prone long reads---a challenging yet unsolved problem.
Last, it is interesting to apply the constructed MISs as the set of representative $k$-mers
in sketching large-scale sequences.

\bibliographystyle{plainurl}
\bibliography{tolerance}

\end{document}